\documentclass[10pt,conference]{IEEEtran}
\usepackage{amsmath,amssymb}
\usepackage{graphicx}
\usepackage{mathrsfs}
\usepackage{tikz}
\usepackage{booktabs}
\usepackage{floatflt}
\usepackage{enumerate}
\usepackage{psfrag}	
\usepackage{array}
\usepackage{multirow,hhline}
\usepackage{exscale}
\usepackage{color}
\usepackage{colortbl}
\usepackage{epsfig}
\usepackage{cite}
\usepackage{amsthm}

\newtheorem{theorem}{Theorem}
\newtheorem{lemma}{Lemma}

\newtheorem{definition}{Definition}

\begin{document}



\IEEEoverridecommandlockouts
\title{The Capacity Region of the 3-User Gaussian Interference Channel with Mixed Strong-Very Strong Interference}
\author{
\IEEEauthorblockN{Anas Chaaban and Aydin Sezgin}
\IEEEauthorblockA{Emmy-Noether Research Group on Wireless Networks\\
Institute of Telecommunications and Applied Information Theory\\
Ulm University, 89081 Ulm, Germany\\
Email: {anas.chaaban@uni-ulm.de, aydin.sezgin@uni-ulm.de}}
\thanks{%
This work is supported by the German Research Foundation, Deutsche
Forschungsgemeinschaft (DFG), Germany, under grant SE 1697/3.%
}
}

\maketitle


\begin{abstract}
We consider the 3-user Gaussian interference channel and provide an outer bound on its capacity region. Under some conditions, which we call the mixed strong-very strong interference conditions, this outer bound is achievable. These conditions correspond to the case where at each receiver, one transmitter is causing strong interference and the other is causing very strong interference. Therefore, we characterize the capacity region of the 3-user interference channel with mixed strong-very strong interference.
\end{abstract}


\section{Introduction}
The interference channel (IC) is a problem in information theory which is being studied since decades. The simplest form of an IC is the 2-user IC where 2 transmit-receive pairs communicate using the same medium and each pair disturbs the other pair with interference. Although this setup is being studied since a long time, its capacity is only known for some cases. In \cite{Carleial_vsi} for instance, the capacity region of the  2-user Gaussian IC was given for the regime that was later called the very strong interference regime. In \cite{Sato}, the capacity region of the 2-user Gaussian IC was obtained for the strong interference regime where each signal arrives to its undesired receiver (where it causes interference) at a higher power than at its desired receiver. Only recently, the sum capacity of the 2-user Gaussian IC was obtained for a noisy interference regime \cite{ShangKramerChen,AnnapureddyVeeravalli,MotahariKhandani}. It was shown that a regime exists where treating interference as noise at each receiver achieves the sum capacity of the IC, and hence the name "noisy interference regime". In general, the capacity of the IC is known within a gap of one bit \cite{EtkinTseWang}.

In addition to the Gaussian IC, the deterministic IC has also been an active research topic recently, which has played an important role in developing new results and gaining new insights into the Gaussian IC. For instance, the deterministic 2-user IC was studied in \cite{BreslerTse} where constant-gap capacity results were obtained, and the cyclically symmetric deterministic K-user interference channel was studied in \cite{BandemerVilarElGamal} where sum capacity results were obtained.

Some extensions of capacity results to Gaussian IC's with more users exist. For instance, in \cite{SridharanJafarianVishwanathJafar} it was shown that using lattice codes can enlarge the very strong interference regime of the K-user IC. In \cite{SridharanJafarianVishwanathJafarShamai}, a layered lattice coding scheme for the 3-user IC was proposed, which achieves more than one degree of freedom (DoF), and which was shown to outperform the Han-Kobayashi scheme. In \cite{ShangKramerChen_KUserIC}, a noisy interference regime of the K-user IC was derived as an extension of the the 2-user result. 

The problem of the K-user IC was also considered from a DoF point of view to obtain approximate capacity results. In \cite{CadambeJafar_KUserIC}, interference alignment was used to obtain the DoF of the K-user IC showing that the K-user IC has K/2 DoF. In \cite{JafarVishwanath} the generalized DoF of the symmetric K-user IC were characterized. Approximate capacity were obtained for the many-to-one IC in \cite{BreslerParekhTse} and for the cyclic IC in \cite{ZhouYu}.

In this paper, we derive the capacity region of the 3-user Gaussian IC with what we call "mixed strong-very strong interference". With mixed strong-very strong interference we mean that at each receiver, one of the interfering signals is strong and the other interfering signal is very strong. In this regime, a capacity outer bound is obtained by using a similar approach as Sato's approach in \cite{Sato} for obtaining the capacity of the 2-user IC with strong interference. If the 3-user IC has strong interference, then the rate pair $(R_i,R_j)$ must lie in the capacity region of the multiple access channel (MAC) from transmitters $i$ and $j$ to receiver $i$. This gives us 6 MAC-like bounds. Now if one of the interferers is very strong, i.e. it can be decoded at its undesired receiver (where it is causing very strong interference) while treating the remaining signals as noise, without imposing any additional constraint on the achievable rates, then 3 of the 6 MAC-like bounds can be dropped. The remaining 3 bounds define a rate region that can be achieved by decoding the remaining signals (after subtracting the very strong interference) in a MAC fashion. The conditions for this mixed strong-very strong interference regime are given and the capacity region is characterized.

The rest of the paper is arranged as follows. In Section \ref{Model}, we describe the system model of the 3-user Gaussian IC. The Main results are given in Section \ref{MainResults}. We conclude with Section \ref{Conclusion}.

\section{System Model}
\label{Model}
Consider a 3-user Gaussian IC with the following input-output equations
\begin{align*}
Y_j=\sum_{i=1}^3h_{ij}X_i+Z_j,
\end{align*}
where $h_{ij}\in\mathbb{R}$ is the channel coefficient from transmitter $i$ to receiver $j$, $X_i$ is the transmit signal of transmitter $i$ which has a power constraint $P$
$$\mathbb{E}[X_i^2]\leq P$$ and $Z_j$ is a Gaussian noise with zero mean and unit variance $Z_j\sim\mathcal{N}(0,1)$.

\begin{figure}[ht]
\centering
\includegraphics[width=0.8\columnwidth]{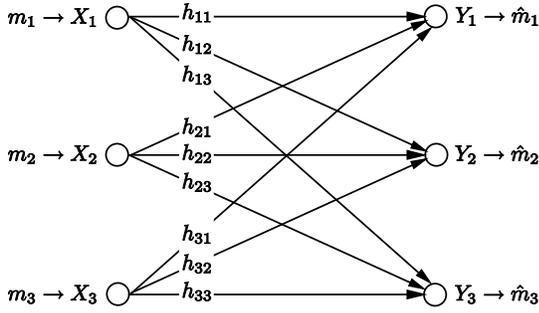}
\caption{A 3-user interference channel.}
\end{figure}

An $(n,2^{nR_1},2^{nR_2},2^{nR_3})$ code for the 3-user IC consists of message sets $$\mathcal{M}_i=\{1,\dots,2^{nR_i}\},$$
encoding functions 
$$f_i:\mathcal{M}_i\to \mathbb{R}^n,$$
and decoding functions
$$g_i:\mathbb{R}^n\to \mathcal{M}_i,$$
$\forall i\in\{1,2,3\}$. A rate tuple $(R_1,R_2,R_3)$ is said to be achievable if there exist a sequence of $(n,2^{nR_1},2^{nR_2},2^{nR_3})$ codes such that the average decoding error probability goes to zero as $n$ increases. The closure of the set of all achievable rate tuples is the capacity region of the 3-user IC and is denoted by $\mathcal{C}$. 

In the following section, we provide an outer bound and an inner bound on $\mathcal{C}$ that coincide if the IC has mixed strong-very strong interference.

\section{Main Result}
\label{MainResults}
Before proceeding to the main result, we will give an example which illustrates the idea. We are going to need the following definition and lemma.
\begin{definition}
\label{Def}
A multiple access channel (MAC) from the set of transmitters $\mathcal{S}\subseteq\{1,2,3\}$ to receiver $j\in\{1,2,3\}$ is denoted by $M(\mathcal{S},j)$ and its corresponding capacity region is denoted by $\mathcal{C}^M(\mathcal{S},j)$. This region is given by
\begin{equation}
\mathcal{C}^M(\mathcal{S},j)=\left\{
\begin{array}{l}
R_i\geq0\ \forall i\in\mathcal{S}: \forall \mathcal{T}\subseteq\mathcal{S}\\
\sum_{i\in\mathcal{T}}R_i\leq\frac{1}{2}\log\left(1+\sum_{i\in\mathcal{T}}h_{ij}^2P_i\right)
\end{array}
\right\}.
\end{equation}
\end{definition}

Now, we are ready to state the following lemma, which provides an outer bound on the capacity region of the 3-user Gaussian IC.
\begin{lemma}
\label{SILemma}
The capacity region of the 3-user Gaussian IC is outer bounded by $\overline{\mathcal{C}}$, i.e.
$$\mathcal{C}\subseteq\overline{\mathcal{C}}$$
where
\begin{align}
\overline{\mathcal{C}}\triangleq\left\{
\begin{array}{l}
(R_1,R_2,R_3)\in\mathbb{R}^3_+: \\
\forall i,j\in\{1,2,3\},\ i\neq j\\
R_i\leq\frac{1}{2}\log(1+h_{ii}^2P)\\
(R_i,R_j)\in\mathcal{C}^M(\{i,j\},j)\text{ if } h_{ij}^2\geq h_{ii}^2
\end{array}\right\}
\end{align}
\end{lemma}
\begin{proof}
The proof uses similar arguments as in the 2-user Gaussian IC with strong interference \cite{Sato}. Details are given in Appendix \ref{SI}.
\end{proof}

If $h_{ij}^2\geq h_{ii}^2$, then interference from transmitter $i$ to receiver $j$ is said to be strong. Lemma \ref{SILemma} states that if transmitter $i$ causes strong interference to receiver $j$, then any achievable rate pair $(R_i,R_j)$ must lie within the capacity region of the MAC $M(\{i,j\},j)$. Thus, we have an outer bound on $\mathcal{C}$. Next, we examine a case where this outer bound simplifies to an achievable inner bound, thus obtaining $\mathcal{C}$ for this case.

Consider a 3-user Gaussian IC where 
\begin{align*}
h_{12}^2,h_{13}^2&\geq h_{11}^2,\\
h_{21}^2,h_{23}^2&\geq h_{22}^2,\\
h_{31}^2,h_{32}^2&\geq h_{33}^2.
\end{align*}
That is, the interference caused by each transmitter at each receiver is strong. In this case, according to Lemma \ref{SILemma} the capacity region of this IC is outer bounded by
\begin{align}
\overline{\mathcal{C}}_{si}=\left\{
\begin{array}{l}
(R_1,R_2,R_3)\in\mathbb{R}^3_+:\\
\forall i,j\in\{1,2,3\},\ i\neq j\\
(R_i,R_j)\in\mathcal{C}^M(\{i,j\},j)
\end{array}\right\}
\end{align}
Notice that in $\overline{\mathcal{C}}_{si}$, we dropped the bound $R_i\leq\frac{1}{2}\log(1+h_{ii}^2P)$ since it is already included in the bound $(R_i,R_j)\in\mathcal{C}^M(\{i,j\},j)$.

Moreover, assume as an example that one interference link is very strong at each receiver, say $h_{31}$, $h_{12}$, $h_{23}$. That is, in addition to 
\begin{align}
\label{VSI2}
h_{21}^2&\geq h_{22}^2,\\
h_{32}^2&\geq h_{33}^2,\\
h_{13}^2&\geq h_{11}^2,
\end{align}
the IC satisfies the following conditions
\begin{align}
\label{VSI1}
h_{31}^2&\geq h_{33}^2\left(1+h_{11}^2P+h_{21}^2P\right),\\
h_{12}^2&\geq h_{11}^2\left(1+h_{22}^2P+h_{32}^2P\right),\\
\label{VSI3}
h_{23}^2&\geq h_{22}^2\left(1+h_{13}^2P+h_{33}^2P\right).
\end{align}

We call these conditions the 'mixed-strong very strong' interference conditions since at each receiver we have one strong and one very strong interferer according to inequalities (\ref{VSI2})-(\ref{VSI3}). Under these condition, it can be shown that the following bounds
\begin{align*}
(R_1,R_2)\in\mathcal{C}^M(\{1,2\},2),\\
(R_2,R_3)\in\mathcal{C}^{M}(\{2,3\},3),\\
(R_1,R_3)\in\mathcal{C}^{M}(\{1,3\},1),
\end{align*}
in $\overline{\mathcal{C}}_{si}$ are redundant and can be dropped out. The details showing why these bounds are redundant are given in Appendix \ref{Mixed-SI-VSI}. Therefore, the capacity region $\mathcal{C}$ is outer bounded as follows if we have mixed strong-very strong interference
\begin{align}
\mathcal{C}\subseteq\overline{\mathcal{C}}_{si}\triangleq\left\{
\begin{array}{l}
(R_1,R_2,R_3)\in\mathbb{R}^3_+:\\
(R_1,R_2)\in\mathcal{C}^M(\{1,2\},1)\\
(R_2,R_3)\in\mathcal{C}^{M}(\{2,3\},2)\\
(R_1,R_3)\in\mathcal{C}^{M}(\{1,3\},3)
\end{array}\right\}
\end{align}

Now consider the following achievable scheme for the 3-user Gaussian IC satisfying conditions (\ref{VSI2})-(\ref{VSI3}). Transmitter $i$ uses a Gaussian codebook to encode its message $m_i$ into an i.i.d. Gaussian sequence $X_i^n$ where $X_i\sim\mathcal{N}(0,P)$, $i\in\{1,2,3\}$. Each receiver decodes the very strong interferer first while treating the remaining signals as noise. Then it subtracts the contribution of the very strong interferer and decodes the remaining two signals in a MAC fashion. So at receiver 1, the very strong interference signal $X_3^n$ is decoded first, while treating $X_1^n$ and $X_2^n$ as noise. This is possible with arbitrarily small error probability if
\begin{align}
\label{TIN1}
R_3\leq\frac{1}{2}\log\left(1+\frac{h_{31}^2P}{1+h_{11}^2P+h_{21}^2P}\right).
\end{align}
Then, receiver 1 subtracts $h_{31}X_3^n$ from $Y_1^n$. Receiver 1 is able then to decode $X_1^n$ and $X_2^n$ with a low error probability if
\begin{align}
\label{MAC1}
(R_1,R_2)\in\mathcal{C}^M(\{1,2\},1).
\end{align}
A similar procedure is done at the other receivers, resulting in the following rate constraints
\begin{align}
\label{TIN2}
&R_1\leq\frac{1}{2}\log\left(1+\frac{h_{12}^2P}{1+h_{22}^2P+h_{32}^2P}\right),\\
\label{TIN3}
&R_2\leq\frac{1}{2}\log\left(1+\frac{h_{23}^2P}{1+h_{33}^2P+h_{13}^2P}\right),\\
\label{MAC2}
&(R_2,R_3)\in\mathcal{C}^M(\{2,3\},2),\\
\label{MAC3}
&(R_1,R_3)\in\mathcal{C}^M(\{1,3\},3).
\end{align}

Choose the rates of the messages to satisfy (\ref{MAC1}), (\ref{MAC2}), and (\ref{MAC3}). If the IC satisfies conditions (\ref{VSI2})-(\ref{VSI3}), then the rate constraints (\ref{TIN1}), (\ref{TIN2}), and (\ref{TIN3}) are automatically satisfied by the choice of the rates according to (\ref{MAC1}), (\ref{MAC2}), and (\ref{MAC3}). Thus, the receivers are able to decode their messages with low error probability if
\begin{align}
(R_1,R_2,R_3)\in\underline{\mathcal{C}}\triangleq\left\{
\begin{array}{l}
(R_1,R_2,R_3)\in\mathbb{R}^3_+:\\
(R_1,R_2)\in\mathcal{C}^M(\{1,2\},1)\\
(R_2,R_3)\in\mathcal{C}^{M}(\{2,3\},2)\\
(R_1,R_3)\in\mathcal{C}^{M}(\{1,3\},3)
\end{array}\right\}
\end{align}

\begin{figure}
\centering
\psfragscanon
\psfrag{x}[]{$R_1$}
\psfrag{y}[]{$R_2$}
\psfrag{z}[]{$R_3$}
\includegraphics[width=0.8\columnwidth]{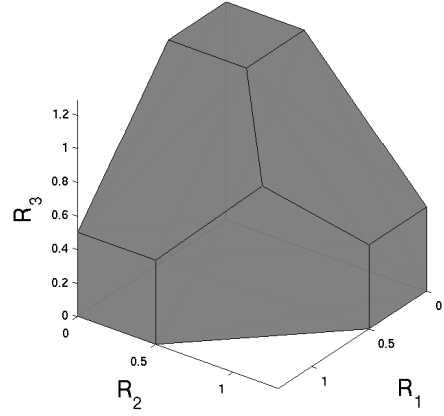}
\caption{The capacity region of a 3-user Gaussian IC with: $P=5$, $h_{11}=h_{22}=h_{33}=1$, $h_{21}=h_{32}=h_{13}=1.1$, $h_{31}=h_{12}=h_{23}=4$}
\label{CR}
\end{figure}

Since in this case $\underline{\mathcal{C}}=\overline{\mathcal{C}}_{si}$, the capacity region of the 3-user Gaussian IC with conditions (\ref{VSI2})-(\ref{VSI3}) is given by
$$\mathcal{C}=\underline{\mathcal{C}}=\overline{\mathcal{C}}_{si}.$$
The capacity region of an example 3-user Gaussian IC with mixed strong-very strong interference is shown in Figure \ref{CR}.

In the previous derivation, we have assumed that $h_{31}$, $h_{12}$, and $h_{23}$ are the very strong interference links while the other links are strong. Same holds if one element of each of these sets $\{h_{21},h_{31}\}$, $\{h_{12},h_{32}\}$, and $\{h_{13},h_{23}\}$ is the very strong interference link and the remaining three links are strong. In general, it is required that one interferer is strong and the other is very strong at each receiver. The following theorem collects all cases in one expression.

\begin{theorem}
\label{Mixed-Thm}
In a 3-user Gaussian IC, if one transmitter $j_1$ causes very strong interference to receiver $j$, and the other transmitter $j_2$ causes strong interference, i.e. $\forall j\in\{1,2,3\}$
\begin{align*}
h_{j_1j}^2&\geq h_{j_1j_1}^2\left(1+h_{jj}^2P+h_{j_2j}^2P\right),\\
h_{j_2j}^2&\geq h_{j_2j_2}^2,
\end{align*}
where $\{j_1,j_2,j\}=\{1,2,3\}$, then its capacity region is given by
\begin{align*}
\mathcal{C}=\left\{
\begin{array}{l}
(R_1,R_2,R_3)\in\mathbb{R}_+^3:\\
\forall j\in\{1,2,3\}\\
(R_j,R_{j_2})\in\mathcal{C}^M(\{j,j_2\},j)
\end{array}\right\}.
\end{align*}
\end{theorem}
\begin{proof}
The converse is given in Appendix \ref{Mixed-SI-VSI}. The achievability is similar to the scheme discussed above for the considered IC, but is also given in Appendix \ref{Mixed-SI-VSI} for the general case for the sake of completeness.
\end{proof}

\section{Conclusion}
\label{Conclusion}
We considered the 3-user Gaussian IC and obtained an outer bound for its capacity region. We have shown that this outer is tight if the IC has mixed strong-very strong interference. That is, if one interferer is strong and the other is very strong at each receiver of the IC. In this case, each receiver can start by decoding the very strong interferer while treating the other 2 signals as noise, then subtract its contribution and decode the remaining 2 signals in a MAC fashion. Thus, the capacity region in this case is characterized.

\section{Acknowledgment}
The authors would like to thank Bernd Bandemer (Stanford University) for fruitful discussions.

\bibliography{/home/chaaban/tex/myBib}

\begin{appendices}

\section{Proof of Lemma \ref{SILemma}}
\label{SI}
Clearly, the achievable rates must satisfy the single user bounds
\begin{align*}
R_1&\leq\frac{1}{2}\log(1+h_{11}^2P)\\
R_2&\leq\frac{1}{2}\log(1+h_{22}^2P)\\
R_3&\leq\frac{1}{2}\log(1+h_{33}^2P).
\end{align*}

Now we derive the other bounds. A genie gives the first receiver $X_3^n$ as additional information. The obtained genie aided channel has a larger capacity region than the original IC. Now consider a rate tuple $(R_1,R_2,R_3)$ in the capacity region of this genie aided channel. This means that receiver $i$ is able to decode $X_i^n$ reliably, $i\in\{1,2,3\}$. Since the first receiver is able to decode $X_1^n$, and since it knows $X_3^n$, then it is able to construct 
\begin{align*}
\tilde{Y}_2^n&=\frac{h_{22}}{h_{21}}(Y_1^n-h_{11}X_1^n-h_{31}X_{3}^n)+h_{32}X_3^n+h_{12}X_1^n\\
&=h_{12}X_1^n+h_{22}X_2^n+h_{32}X_3^n+\frac{h_{22}}{h_{21}}Z_1^n
\end{align*}
If $h_{21}^2\geq h_{22}^2$, then $\tilde{Y}_2^n$ is a less noisy version of $Y_2^n$. So if the second receiver is able to decode $X_2^n$ then so does the first receiver. Thus
$(R_1,R_2)$ is contained in the capacity region of the MAC $M(\{1,2\},1)$, i.e. $$(R_1,R_2)\in\mathcal{C}^{M}(\{1,2\},1)  \text{ if } h_{21}^2\geq h_{22}^2.$$
The other bounds can be derived similarly.

\section{Proof of Theorem \ref{Mixed-Thm}}
\label{Mixed-SI-VSI}

Assume that one interferer is very strong and the other interferer is strong at each receiver. That is, at receiver $j$
\begin{align}
\label{VSI-Cond}
h_{j_1j}^2&\geq h_{j_1j_1}^2\left(1+h_{jj}^2P+h_{j_2j}^2P\right)\geq h_{j_1j_1}^2,\\
\label{SI-Cond}
h_{j_2j}^2&\geq h_{j_2j_2}^2,
\end{align}
where $\{j_1,j_2,j\}=\{1,2,3\}$ and where we denoted the strong interferer at receiver $j$ by $j_2$ and the very strong interferer by $j_1$. Notice that the very strong interference also satisfies the strong interference condition $h_{j_1j}^2\geq h_{j_1j_1}^2$. Thus, according to Lemma \ref{SILemma}, the capacity region of the given 3-user Gaussian IC is contained in the following set
\begin{align}
\label{Region}
\mathcal{C}\subseteq\left\{
\begin{array}{l}
(R_1,R_2,R_3)\in\mathbb{R}^3_+:\\
R_1\leq\frac{1}{2}\log(1+h_{11}^2P)\\
R_2\leq\frac{1}{2}\log(1+h_{22}^2P)\\
R_3\leq\frac{1}{2}\log(1+h_{33}^2P)\\
(R_1,R_2)\in\mathcal{C}^M(\{1,2\},1)\\
(R_1,R_2)\in\mathcal{C}^M(\{1,2\},2)\\
(R_2,R_3)\in\mathcal{C}^{M}(\{2,3\},2)\\
(R_2,R_3)\in\mathcal{C}^{M}(\{2,3\},3)\\
(R_1,R_3)\in\mathcal{C}^{M}(\{1,3\},1)\\
(R_1,R_3)\in\mathcal{C}^{M}(\{1,3\},3)
\end{array}\right\}.
\end{align}
We will show that if $h_{j_1j}$ satisfies (\ref{VSI-Cond}), then the bound $$(R_{j_1},R_j)\in\mathcal{C}^M(\{j_1,j\},j)$$ is redundant and can be dropped out. To show this, we expand this bound by using Definition \ref{Def} as follows
\begin{align}
(R_{j_1},R_{j})&\in\mathcal{C}^M(\{j_1,j\},j) \Leftrightarrow\nonumber\\
\label{B1}
R_{j_1}&\leq \frac{1}{2}\log(1+h_{j_1j}^2P)\\
\label{B2}
R_{j}&\leq \frac{1}{2}\log(1+h_{jj}^2P)\\
\label{B3}
R_{j_1}+R_{j}&\leq \frac{1}{2}\log(1+h_{j_1j}^2P+h_{jj}^2P).
\end{align}
From condition (\ref{VSI-Cond}) and the region (\ref{Region}), we have
\begin{align*}
R_{j_1}&\leq \frac{1}{2}\log(1+h_{j_1j_1}^2P)\\
&\leq \frac{1}{2}\log(1+h_{j_1j}^2P).
\end{align*}
Thus, (\ref{B1}) is redundant. The bound (\ref{B2}) is already included in the single user bounds in (\ref{Region}) and thus is also redundant. The last bound (\ref{B3}) can be shown to be redundant as follows
\begin{align*}
R_{j_1}+R_{j}&\leq\frac{1}{2}\log(1+h_{j_1j_1}^2P)+\frac{1}{2}\log(1+h_{jj}^2P)\\
&\stackrel{(a)}{\leq}\frac{1}{2}\log\left(1+\frac{h_{j_1j}^2P}{1+h_{jj}^2P+h_{j_2j}^2P}\right)\\
&\ \ \ +\frac{1}{2}\log(1+h_{jj}^2P)\\
&\leq\frac{1}{2}\log\left(1+\frac{h_{j_1j}^2P}{1+h_{jj}^2P}\right)+\frac{1}{2}\log(1+h_{jj}^2P)\\
&=\frac{1}{2}\log\left(1+h_{j_1j}^2P+h_{jj}^2P\right).
\end{align*}
where in $(a)$ we used (\ref{VSI-Cond}). Thus, since all three bounds (\ref{B1}), (\ref{B2}), and (\ref{B3}) are redundant, then $(R_{j_1},R_{j})\in\mathcal{C}^M(\{j_1,j\},j)$ is redundant and can be dropped out. This can be shown for $j=1,2,3$. Therefore, at each receiver, only one MAC bound remains and we can write
\begin{align}
\label{Region2}
\mathcal{C}\subseteq\left\{
\begin{array}{l}
(R_1,R_2,R_3)\in\mathbb{R}_+^3:\\
\forall j\in\{1,2,3\}\\
(R_j,R_{j_2})\in\mathcal{C}^M(\{j,j_2\},j)
\end{array}\right\},
\end{align}
where $j_2$ is the strong interferer at receiver $j$. 

But this region, given by (\ref{Region2}) is achievable as follows. Each transmitter encodes its message $m_i$ into an i.i.d. Gaussian codeword of length $n$, $X_i^n$ where $X_i\sim\mathcal{N}(0,P)$. 

At each receiver, the very strong interference is decoded first while treating all the other signals as noise. So at receiver $j$, the very strong interferer $X_{j_1}^n$ is decoded first with effective noise given by $h_{jj}X_j^n+h_{j_2j}X_{j_2}^n+Z_j^n$. The decoding probability of error can be made arbitrarily small if
\begin{equation}
\label{Constraint1}
R_{j_1}\leq\frac{1}{2}\log\left(1+\frac{h_{j_1j}^2P}{1+h_{jj}^2P+h_{j_2j}^2P}\right).
\end{equation}
Then, $h_{j_1j}X_{j_1}^n$ is subtracted, and $X_{j_2}^n$ and $X_j^n$ are decoded in a MAC fashion. Thus, for reliable decoding their rates must satisfy
\begin{equation}
\label{Constaint2}
(R_{j_2},R_{j})\in\mathcal{C}^M(\{j_2,j\},j).
\end{equation}
This is done at each receiver. The MAC rate constraint at receiver $j_1$ requires
\begin{align}
\label{Constraint3}
R_{j_1}\leq\frac{1}{2}\log(1+h_{j_1j_1}^2P).
\end{align}
If condition (\ref{VSI-Cond}) is satisfied, then the constraint (\ref{Constraint1}) is less binding than (\ref{Constraint3}), and thus (\ref{Constraint1}) is redundant. As a result, every rate pair given by
\begin{equation*}
(R_{j_2},R_{j})\in\mathcal{C}^M(\{j_2,j\},j).
\end{equation*}
is achievable. Consequently, if
\begin{align*}
h_{j_1j}^2&\geq h_{j_1j_1}^2\left(1+h_{jj}^2P+h_{j_2j}^2P\right),\\
h_{j_2j}^2&\geq h_{j_2j_2}^2,
\end{align*}
$\forall j\in\{1,2,3\}$ where $\{j_1,j_2,j\}=\{1,2,3\}$, then the capacity region of the 3-user Gaussian IC is
\begin{align*}
\mathcal{C}=\left\{
\begin{array}{l}
(R_1,R_2,R_3)\in\mathbb{R}_+^3:\\
\forall j\in\{1,2,3\}\\
(R_j,R_{j_2})\in\mathcal{C}^M(\{j,j_2\},j)
\end{array}\right\}.
\end{align*}

\end{appendices}

\end{document}